\RequirePackage{fix-cm}
\documentclass{ws-ijmpd}

\RequirePackage{graphicx}

\RequirePackage{flushend}
\RequirePackage{mathptmx}      
\RequirePackage{latexsym}
\RequirePackage[numbers,sort&compress]{natbib}
\RequirePackage[colorlinks,citecolor=blue,urlcolor=blue,linkcolor=blue]{hyperref}
\usepackage{hyperref}
\usepackage{mathtools}
\usepackage{amsmath}
\usepackage{cuted}
\usepackage{mathtools,times,cuted}

\newcommand{\hunit}{$\rm{km \ sec^{-1} \ Mpc^{-1}}$}

 \usepackage[latin1]{inputenc}
\usepackage{color}
\usepackage{url}
\usepackage{graphicx}
\hypersetup{linkcolor=red,citecolor=blue,filecolor=cyan,urlcolor=blue}
\usepackage{float}
\usepackage{amsmath}
\usepackage{tikz}
\usepackage{graphicx}
\usepackage{subfigure}

\usepackage{latexsym,color}
\usepackage{amssymb}
\usepackage{xcolor}
\usepackage{amsfonts,dsfont}
\usepackage{amssymb,bm,dcolumn,booktabs,threeparttable,multirow,fleqn}
\usepackage[T1]{fontenc}
\usepackage{ae,aecompl,latexsym}
\def\be{\begin{equation}}
\def\ee{\end{equation}}
\def\bea{\begin{eqnarray}}

\def\eea{\end{eqnarray}}

\usepackage{physics}

\usepackage[varg]{txfonts}
\begin{document}

\markboth{}
{Constraints on the Transition Redshift using  Hubble Phase Space Portrait}

%
\catchline{}{}{}{}{}
%







\title{Constraints on the Transition Redshift using  Hubble Phase Space Portrait
}


\author{Darshan Kumar$^1$, 
        Deepak Jain$^2$,
        Shobhit Mahajan$^1$,
        Amitabha Mukherjee$^1$\and
        Akshay Rana$^3$ 
}



\address{$^1$Department of Physics and Astrophysics, University of Delhi,
Delhi 110007, India \\
$^2$Deen Dayal Upadhyaya College, University of Delhi,
Dwarka, New Delhi 110078, India \\
$^3$St. Stephen's College, University of Delhi,
Delhi 110007, India 
}

\date{Received: date / Accepted: date}

\maketitle

\begin{abstract}
One of the most significant discoveries in modern cosmology is that the universe is currently in a phase of accelerated expansion after a switch from a decelerated expansion. The redshift corresponding to this epoch is commonly referred to as the transition redshift $z_t$. In this work we put constraints on the transition redshift with both model-independent and model-dependent approaches. We consider the  recently compiled database of 32 Hubble parameter measurements and the Pantheon sample of Type Ia Supernovae (SNe). In order to include the possible systematic effects in this analysis, we use the full covariance matrix of systematic uncertainties for the Hubble parameter measurements. We plot a Hubble Phase Space Portrait (HPSP) between $\dot{H}(z)$ and $H(z)$ in a model-independent way. From this HPSP diagram, we estimate the transition redshift as well as the current value of the equation of state parameter $\omega_0$ in a model-independent way. By considering H(z) measurements, we find the best fit value of $z_t=0.591^{+0.332}_{-0.332}$ and $\omega_0=-0.677^{+0.238}_{-0.238}$. We obtain the best fit value of $z_t=0.849^{+0.117}_{-0.117}$ and $\omega_0=-0.870^{+0.013}_{-0.013}$ using the Pantheon database. Further, we also use a model dependent approach to determine $z_t$. Here, we consider a non-flat $\Lambda$CDM model as a background cosmological model. We reconstruct the cosmic triangle plot among $\log(\Omega_{m0})$, $-\log(2\Omega_{\Lambda0})$ and $3\log(1+z_t)$ where the constraints of each parameter are determined by the location in this triangle plot. Using $\Omega_{m0}$ and $\Omega_{\Lambda0}$ values, we find the best value of the transition redshift $z_t=0.619^{+0.580}_{-0.758}$, which is in good agreement with the Planck 2018 results at $1\sigma$ confidence level. We also simulate the observed Hubble parameter measurements in the redshift range $0<z<2$ and perform the same analysis to estimate the transition redshift.

\end{abstract}

\section{Introduction}
It is now widely believed that we live in a phase of accelerated expansion of the universe, which has been corroborated by several independent observations. The most direct and obvious evidence of the present accelerating stage of the universe comes from the luminosity distance versus redshift relation measurements using  Type Ia supernovae (SNe) which are confirmed by two different research groups \cite{Riess1998,Perlmutter1999}. Over the decades, the number of SNe Type Ia catalogues have grown substantially. Even now, the SNe Type Ia observations provide the most  direct evidence for the current cosmic acceleration  (\cite{2006A&A...447...31A,2007ApJ...659...98R,2007ApJ...666..716D,2008ApJ...686..749K,2010ApJ...716..712A,2012ApJ...746...85S,Goober2018}). This discovery is also supported by complementary observations such as the  Cosmic Microwave Background (CMB) radiation measurements (\cite{2011ApJS..192...18K,2011ApJS..192...16L,Planck2018}), Baryonic Acoustic Oscillations (BAO) measurements (\cite{2005ApJ...633..560E,2007MNRAS.381.1053P,2011AJ....142...72E,2013AJ....145...10D,2019JCAP...10..044C}) and Hubble parameter (H(z)) measurements (\cite{Stern_2010,2013ApJ...764..138F,2013ApJ...766L...7F,10.1093/mnrasl/slv037,Moresco_2016,2017ApJ...835...26F}). From a theoretical point of view, the simplest and most popular $\Lambda$CDM model supports such an  accelerating phase, with free parameters being tightly constrained by  Planck Collaboration et al. (2020) \cite{Planck2018}. In the standard model of cosmology, there was an era of decelerated expansion due to matter (\cite{2008PASP..120..235R,2012CRPhy..13..566M,2018RPPh...81a6901H}).

 Given the transition from a decelerated to an accelerated expansion of the  universe, it is important  to precisely determine the  redshift at which the universe started to accelerate from a decelerated phase. The parameter corresponding to such a transition from decelerated to an accelerated phase  is referred  as the transition redshift $z_t$ and it  may be treated as a new cosmic parameter along with the present deceleration parameter $q_0$ and the Hubble constant $H_0$. It is now commonly accepted that the upper bound on the transition redshift is less than unity, which has been confirmed by multiple independent investigations in both model-dependent and model-independent approaches (\cite{2013ApJ...764..138F,2013ApJ...766L...7F,2014JCAP...10..057S,2006ApJ...649..563S,2008MNRAS.390..210C,2015JCAP...12..045R,2012arXiv1205.4688L,Moresco_2016,2018JCAP...05..073J,velasquez2020observational,2022MNRAS.509.5399C}). Shapiro and Turner (2006) in their analysis adopted a model-dependent approach and using SNe Type Ia observations, they put a constraint on the deceleration parameter $q(z)$ and transition redshift $z_t$ \cite{2006ApJ...649..563S} . By considering  different dynamical dark energy models, A. Melchiorri et al. (2007) \cite{2007PhRvD..76d1301M} put constraints on  $z_t$. Similarly, D. Rapetti et al. (2007) developed a new kinematic method  to estimate $z_t$ using the measurements of X-ray cluster gas mass fraction  \cite{2007MNRAS.375.1510R}. On the other hand, people have also studied the transition redshift in modified gravity models (\cite{2015PhRvD..91l4037C}).

In addition, by considering the dynamical nature of $q(z)$, several authors have explored  constraints on the  transition redshift. For example, Cunha and Lima (2008) have used a linear parametrization of the deceleration parameter i.e. $q(z)=q_{0}+q_{1} z$ \cite{2008MNRAS.390..210C}. In their work, the value of transition redshift is $z_{\mathrm{t}}=0.43_{-0.05}^{+0.09}$ which was obtained by using $182$ SNe Type Ia (\cite{2007ApJ...659...98R}).  By using the SNLS data set, Astier et al. (2006) find the value to be $z_{\mathrm{t}}=0.61_{-0.21}^{+3.68}$ \cite{2006A&A...447...31A}. Rani et al. (2015) also used the linear parametrization of the deceleration parameter and by using a joint analysis of age of galaxies, strong gravitational lensing and SNe Ia data, they found $z_{\mathrm{t}} \approx 0.98$ \cite{2015JCAP...12..045R}. Along the  same lines, other researchers have updated the study by introducing additional data sets and modifying the parametrizations for $q(z)$. For the parametrization $q(z)=q_{0}+q_{1} z /(1+z)$ and using 307 SNe Type Ia together with BAO and $H(z)$ data, Xu et al. (2009) found  $z_{\mathrm{t}}=0.609_{-0.070}^{+0.110}$ \cite{2009JCAP...07..031X}. This type of parametrization has also been investigated by {\O}.  {Elgar{\o}y}, and T. {Multam{\"a}ki} (2006) \cite{2006JCAP...09..002E}. Using different datasets such as Gamma Ray Bursts (GRBs), Hubble Observational Data $H(z)$, BAO, CMB, Galaxy Clusters, lookback time etc., many authors have reconstructed the deceleration parameter  and put constraints on the transition redshift (\cite{2012JCAP...01..018N,2006MNRAS.368..371W,2009JCAP...07..031X,2008MNRAS.390..210C,2007PhRvD..75d3520G,2009arXiv0905.2628L,2013PhLB..726...72F,2013ApJ...766L...7F,2004ApJ...607..665R,2011PhLB..706..116C,2016JCAP...02..066V}). These authors have considered different types of parametrizations for $q(z)$, besides the aforementioned two parametrizations. Recently, by considering $\Lambda$CDM model as a background cosmological model, Planck Collaboration et al. (2020) put a constraint on the transition redshift $z_t=0.632\pm0.018$ \cite{Planck2018}.

The determination of the deceleration parameter $q$ and hence transition redshift $z_t$, as one may see, is an important and relevant subject in modern cosmology along with the determination of the current value of Hubble parameter $H_0$. The determination of these parameters plays a vital role in the understanding of the universe. An alternate and innovative approach for accessing cosmic parameters is a model-independent one  by means of the study of Hubble parameter estimated from the Cosmic Chronometers (CC) approach.

In the literature, a statistical technique of obtaining cosmic parameters has been developed by Seikel et al. (2012) \cite{2012JCAP...06..036S}. This method is called Gaussian Process (GP). In cosmology, using this statistical and non-parametric method, one can reconstruct redshift dependent cosmological functions such as the  expansion rate, luminosity distance etc. Using cosmological observations corresponding to these functions, one can  put constraints on different cosmological parameters. By adopting this model-independent approach, Jesus et al. (2020) found the transition redshift as $z_{t}=0.59_{-0.11}^{+0.12}$ from $H(z)$ measurements \cite{2020JCAP...04..053J}. On the other hand, using the recent and largest up-to-date Pantheon compilation of SNe Type Ia, Scolnic et al. (2018) obtained $z_{t}=0.683_{-0.082}^{+0.110}$ in a non-parametric way \cite{2018ApJ...859..101S}. Further, using the  GP approach, the analysis of Lin et al. (2019) yields $z_t=0.59_{-0.05}^{+0.05}$  \cite{2019ChPhC..43g5101L}. In their analysis, they reconstructed the equation of state of dark energy and cosmic expansion using Gaussian Process from the Pantheon data  and $H(z)$.

In the present work we use the updated Hubble parameter measurements and observations of Type Ia SNe. We obtain the transition redshift in both a model-independent and model-dependent way. Firstly, we carry out the analysis in a \textbf{model-independent way}. With the help of Gaussian Process, we reconstruct the Hubble parameter $H(z)$ and its time-derivative $\dot{H}(z)$ from Hubble parameter measurements and Type Ia SNe observations. We plot a diagram between $\dot{H}(z)$ versus $H(z)$ which is generally referred to as the Hubble Phase Space Portrait (HPSP). By analysing this diagram, we put constraints on  $z_t$ as well as on the  present value of the  equation of state parameter $\omega_0$. By introducing the updated cosmic triangle plot and Hubble Phase Space Portraits, we propose that these representations will help to visualise and estimate the parameters in a unique way. We also simulate the observed Hubble parameter measurements in the redshift range $0<z<2$ to estimate the transition redshift.

 In a \textbf{model-dependent} approach analysis, we consider a non-flat $\Lambda$CDM model as a background cosmological model. After marginalization over $H_0$ using a uniform prior, we put constraints on the cosmological density parameters i.e. $\Omega_{m0}$, $\Omega_{k0}$, and $\Omega_{\Lambda0}$. Using the sum rule of these three cosmological density parameters i.e.  $\Omega_{m0}+\Omega_{k0}+\Omega_{\Lambda0}=1$, we represent them by an equilateral triangle and the  so-called ternary plot, where, each parameter runs parallel to each of the edges on the equilateral triangle. The sum rule must be  satisfied at every  location where lines of constant $\Omega_{m0}$, $\Omega_{k0}$, and $\Omega_{\Lambda0}$ intersect. This is similar to the `cosmic triangle' described by Bahcall et al. (1999) \cite{1999Sci...284.1481B}. Further, using a relation between $z_t$, $\Omega_{m0}$ and $\Omega_{\Lambda0}$, we update the ternary plot and estimate the transition redshift.

The outline of the paper is as follows. In section \ref{obs_dataset},  we discuss the Hubble parameter measurements and Type Ia Supernovae data for determining the comoving  distance and obtaining the Hubble parameter and its derivatives.  Section \ref{background} introduces the background of `Hubble Phase Space Portrait', `Cosmic Triangle' and the methodology used to obtain the constraints on the transition redshift. In section \ref{results}, we discuss the results. Finally, we summarize the  conclusions in Section \ref{discussion}.

\section{Observational Datasets}\label{obs_dataset}
\subsection{Hubble Parameter Measurements}\label{H(z)_dataset}
The Hubble parameter $H(z)$ which reflects the dynamical features of the universe like its expansion rate and evolution history is a crucial parameter in cosmology. It is also useful to investigate the nature of the dark energy since this parameter can be  inferred directly from  astrophysical observations which do not  depend on any underlying cosmological model. There are other probes which do not directly measure $H(z)$, but rather integrated quantities such as luminosity distances. Currently, the most common methods for obtaining $H(z)$ data are (i) using the differential ages of passively evolving galaxies, generally referred to as `Cosmic Chronometers (CC)' \cite{2003ApJ...593..622J,2005PhRvD..71l3001S} (ii) measurements of peaks of acoustic oscillations of baryons (BAO) \cite{2005ApJ...633..560E,2007MNRAS.381.1053P,2009MNRAS.399.1663G,2010deot.book..246B,2009ApJ...691..241B} (iii) Redshift Drift \cite{1962ApJ...136..319S,1962ApJ...136..334M}.

In this work, we consider the $H(z)$ measurements obtained by using Cosmic Chronometers. The Hubble parameter $H(z)$ in terms of redshift $z$ can be expressed as 

\begin{equation}\label{eq_1}
    H(z)=-\dfrac{1}{1+z}\dfrac{ dz}{ dt}
\end{equation}

However, when employing this approach to calculate $H(z)$, significant caution must be exercised in the selection of galaxies. Stars are continually born in early developing galaxies and the emission spectra will be dominated by the young stellar population. As a result, passively developing red galaxies are utilised to properly estimate the differential ageing of the universe since their light is dominated by the elderly star population \cite{2003ApJ...593..622J}. For  measurements of $H(z)$, the term $dt$ is  to be estimated using the differential age evolution of the universe in a given interval of redshift $dz$. For the estimation of $dt$, Moresco et al. (2012) suggested the use of a direct spectroscopic observable (4000\AA$~$break) which is known to be linearly related to the age of the stellar population of a galaxy at fixed metallicity \cite{2012JCAP...08..006M}. Thus, the measurement of $H(z)$, made purely by using spectroscopic observations,  is independent of the cosmological model and has proved to be a strong probe to constrain cosmological models and assumptions. The most recent data compilation of Hubble parameter measurements by Maga\~{n}a et al. (2018) has $31$ measurements of $H(z)$ which are obtained by using the differential ages of passively evolving galaxies \cite{2018MNRAS.476.1036M}. Here we have added one additional datapoint compiled  by Borghi et al. (2022) at $z=0.75$ \cite{2022ApJ...928L...4B}. Thus 32 $H(z)$ measurements used in this analysis are listed in Table \ref{tab_hz}, spanning the redshift range $0.07 \leq z \leq 1.965$. 

\begin{table}
\centering
\tbl{Most recent compilation of updated Hubble parameter measurements.}
{\begin{tabular}{lcc}
\hline
$z$ & $H(z)$ [\hunit] & Reference\\
\hline
0.07 & $69.0\pm19.6$ & \cite{2014RAA....14.1221Z}\\
0.09 & $69.0\pm12.0$ & \cite{2005PhRvD..71l3001S}\\
0.12 & $68.6\pm26.2$ & \cite{2014RAA....14.1221Z}\\
0.17 & $83.0\pm8.0$ & \cite{2005PhRvD..71l3001S}\\
0.179 & $75.0\pm4.0$ & \cite{2012JCAP...08..006M}\\
0.199 & $75.0\pm5.0$ & \cite{2012JCAP...08..006M}\\
0.2 & $72.9\pm29.6$ & \cite{2014RAA....14.1221Z}\\
0.27 & $77.0\pm14.0$ & \cite{2005PhRvD..71l3001S}\\
0.28 & $88.8\pm36.6$ & \cite{2014RAA....14.1221Z}\\
0.352 & $83.0\pm14.0$ & \cite{2012JCAP...08..006M}\\
0.3802 & $83.0\pm13.5$ &  \cite{Moresco_2016}\\
0.4 & $95.0\pm17.0$ & \cite{2005PhRvD..71l3001S}\\
0.4004 & $77.0\pm10.2$ &  \cite{Moresco_2016}\\
0.4247 & $87.1\pm11.2$ &  \cite{Moresco_2016}\\
0.4497 & $92.8\pm12.9$ &  \cite{Moresco_2016}\\
0.47 & $89.0\pm50.0$ & \cite{2017MNRAS.467.3239R}\\
0.4783 & $80.9\pm9.0$ &  \cite{Moresco_2016}\\
0.48 & $97.0\pm62.0$ & \cite{Stern_2010}\\
0.593 & $104.0\pm13.0$ & \cite{2012JCAP...08..006M}\\
0.68 & $92.0\pm8.0$ & \cite{2012JCAP...08..006M}\\
\textbf{0.75} & $\mathbf{98.8\pm33.6}$ & \textbf{\cite{2022ApJ...928L...4B}}\\
0.781 & $105.0\pm12.0$ & \cite{2012JCAP...08..006M}\\
0.875 & $125.0\pm17.0$ & \cite{2012JCAP...08..006M}\\
0.88 & $90.0\pm40.0$ & \cite{Stern_2010}\\
0.9 & $117.0\pm23.0$ & \cite{2005PhRvD..71l3001S}\\
1.037 & $154.0\pm20.0$ & \cite{2012JCAP...08..006M}\\
1.3 & $168.0\pm17.0$ & \cite{2005PhRvD..71l3001S}\\
1.363 & $160.0\pm33.6$ & \cite{10.1093/mnrasl/slv037}\\
1.43 & $177.0\pm18.0$ & \cite{2005PhRvD..71l3001S}\\
1.53 & $140.0\pm14.0$ & \cite{2005PhRvD..71l3001S}\\
1.75 & $202.0\pm40.0$ & \cite{2005PhRvD..71l3001S}\\
1.965 & $186.5\pm50.4$ & \cite{10.1093/mnrasl/slv037}\\
\hline
\end{tabular}}\label{tab_hz}
\end{table}

\subsection{Type Ia Supernovae Measurements} \label{Super dataset}
To estimate luminosity distance, we use the Pantheon dataset of Type Ia Supernovae. This dataset has 1048 Type Ia SNe datapoints in the redshift range $0.01\leq z\leq2.26$ \cite{2018ApJ...859..101S}.  In Type Ia SNe analysis, the standard observable quantity is the distance modulus, $\mu_{SNe}$ which is basically the difference between the apparent and absolute magnitude of the Type Ia SNe. The relation which gives the observable measurement is defined as

\begin{equation}
    \mu_{\mathrm{SNe}}=m_{\mathrm{B}}(z)+\alpha \cdot X_{1}-\beta \cdot \mathcal{C}-M_{\mathrm{B}}.
\end{equation}
where $m_{\mathrm{B}}$ is the rest frame B-band peak magnitude, $M_B$ represents absolute B-band magnitude of a fiducial Type Ia SNe with $X_1 = 0$ and $C = 0$, $X_1$ and $C$ represent  the time stretch of light curve and supernova colour at maximum brightness respectively. For the Pantheon sample, the stretch-luminosity parameter $(\alpha)$ and the colour-luminosity parameter $(\beta)$ are calibrated to zero. Thus we are left with two parameters and hence the distance modulus can be defined as 

\begin{equation}
    \mu_{\mathrm{SNe}}=m_{\mathrm{B}}(z)-M_{\mathrm{B}}.
\end{equation}

Further, once we know the distance modulus, luminosity distance $(d_L)$ and uncertainty in the luminosity distance $(\sigma_{d_L})$ for each Type Ia SNe can be estimated as 

\begin{equation} \label{eq_44}
    d_{\mathrm{L}}(z)=10^{\left(m_{\mathrm{B}}-M_{\mathrm{B}}-25\right) / 5}~~~(\mathrm{Mpc}), \quad \sigma_{d_{\mathrm{L}}}=\frac{\ln (10)}{5} d_{\mathrm{L}} \sigma_{m_{\mathrm{B}}}~~~(\mathrm{Mpc})
\end{equation}

From Eq. \ref{eq_44}, it is clear that once we estimate the value of $M_B$, we can find the luminosity distance at a given redshift. Recent studies seem to indicate that with redshift there is no evolution of luminosity (or absolute magnitude) of Type Ia supernovae \cite{2019A&A...625A..15T,2022JCAP...01..053K}. So, it is generally accepted that the Type Ia supernovae sample is normally distributed with a mean absolute magnitude of $M_B = -19.22$ \cite{2020PhRvD.101j3517B}. Finally using the following relation of luminosity distance and comoving distance, we estimate the comoving distance for each Type Ia SNe as

\begin{equation}
    d_{co}=\dfrac{d_L}{1+z}
\end{equation}

\section{Background and Methodology}\label{background}

\subsection{Hubble Phase Space Portrait (HPSP)}
For a spatially flat, homogeneous and isotropic universe, the Friedmann equations are

\begin{equation}\label{fe_1}
    \left(\dfrac{\dot{a}}{a}\right)^2=\dfrac{8\pi G}{3}\rho_T
\end{equation}

\begin{equation}\label{fe_2}
    2\dfrac{\ddot{a}}{a}+\left(\dfrac{\dot{a}}{a}\right)^2=-8\pi Gp_T
\end{equation}

\begin{figure}
    \centering
    \includegraphics[width=1\columnwidth]{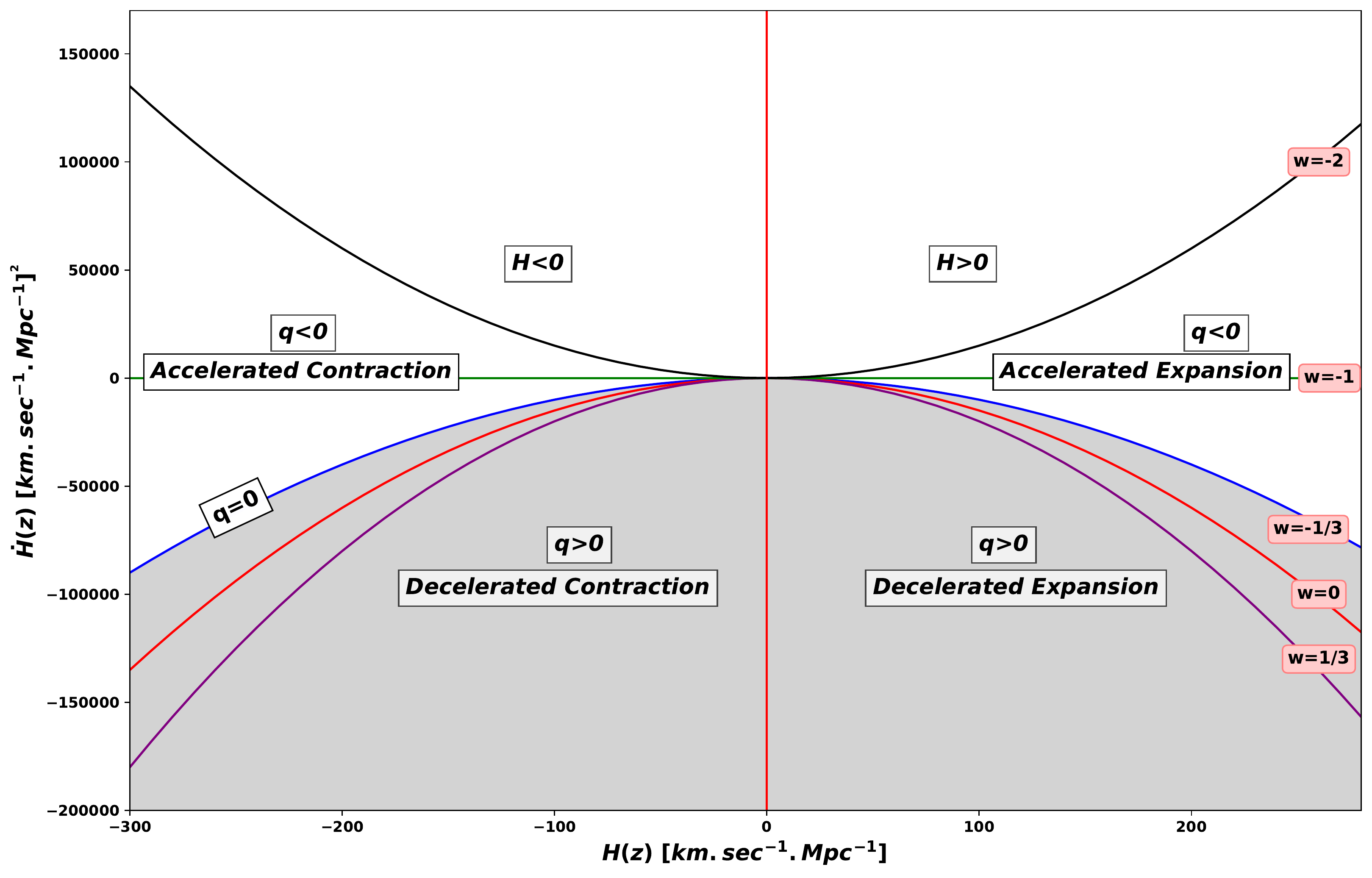}
    \caption{The theoretical reconstruction of Hubble phase space portrait.}
    \label{hpsp_the}
\end{figure}

In these equations, $a(t)$ is scale factor, $\rho_T$ and $p_T$ denote the total energy density and total pressure respectively. By considering $H(t)=\dfrac{\dot{a}}{a}$ and so $\dot{H}+H^2=\dfrac{\ddot{a}}{a}$ and dividing equation (\ref{fe_2}) by equation (\ref{fe_1}), we get

\begin{equation}\label{hpsp_1}
    \dot{H}=-\dfrac{3}{2}(1+\omega)H^2
\end{equation}

where we consider an equation of state parameter $\omega$ which relates the total energy density to total pressure of a barotropic fluid via $p_T=\omega\rho_T$. Depending on the content of the universe,  $\omega$ takes values in the range $-1\leq\omega\leq\dfrac{1}{3}$. Thus we can  reconstruct Friedmann's equations into a single equation i.e. $\dot{H}=\textit{F}(H)$. Using this equation,  by visualizing the trajectories in the $(H,~\dot{H})$ phase space, one can analyze  different cosmic models in a clear and transparent way. Equation (\ref{hpsp_1}) is referred to  as the phase space portrait and its solution is the phase trajectory. For more details about phase space portrait, please refer \cite{2017ChPhC..41l5103E,2018JCAP...02..052A,2019ApJ...871..210E}.  Corresponding to any theory, the phase-space portrait can be drawn in $(H,~\dot{H})$ phase-space. For example, a theoretical reconstruction of the Hubble phase space portrait for different cosmological models is  shown in Figure \ref{hpsp_the}. 

In this framework, we classify the different phases of Figure \ref{hpsp_the} as follows:

\begin{enumerate}
    \item In $(H,~\dot{H})$ phase space, the left half represents a contracting universe since this portion has $H<0$.  On the other hand, the right half  has $H>0$, which represents an expanding universe.
    
    \item For  different values of the equation of state parameter, $\omega$, we can draw lines in the $(H,~\dot{H})$ phase space. For example, for a  matter dominated universe we draw the phase trajectory corresponding to $\omega=0$. Similarly, we plot lines for radiation dominated, curvature dominated and cosmological constant dominated universes corresponding to $\omega=1/3$, $\omega=-1/3$ and $\omega=-1$ respectively.
    
    \item The line corresponding to $w=-1/3$ is generally referred to as ``zero acceleration curve" since the deceleration parameter for this line is zero i.e. $q=0$. This line splits the $(H,~\dot{H})$ phase space into two regions. The upper region represents the accelerated universe where we have $\omega<-1/3$ and $q<0$. On the other hand, the lower region shows the decelerated universe. In this region, the equation of state parameter is $\omega>-1/3$ and deceleration parameter is $q>0$.
    
    \item Therefore, the full $(H,~\dot{H})$ phase space is divided into four different regions. The upper left region represents an \textbf{accelerated contraction} of universe since $H<0$ and $q<0$. The upper right displays an \textbf{accelerated expansion} of universe since $H>0$ and $q<0$. The lower left represents a \textbf{decelerated contraction} of universe where  $H<0$ and $q>0$. And the lower right region represents a \textbf{decelerated expansion} of universe since in this region, we have $H>0$ and $q>0$.
    
\end{enumerate}

As we know, the universe passed from a radiation dominated $(\omega=1/3)$ era  to a matter dominated  $(\omega=0)$ one  in the past. The lower regions characterized the usual FLRW models. As a result we expect that the observations will lie in that region which will show a transition from a decelerated expansion era to an  accelerated expansion era by crossing the zero-acceleration $(q=0)$ line. The value of the Hubble parameter corresponding to this crossing point is referred to  as the transition Hubble parameter value and the corresponding redshift is called the transition redshift. By analyzing the  $(H,~\dot{H})$ phase space, one can easily estimate the transition Hubble parameter value and transition redshift value. Hence, in this analysis, we focus only on this inner right region of the  $(H,~\dot{H})$ phase space. 

\subsection{The Cosmic Triangle}\label{cosmic_triangle}
A triangle or ternary diagram is a graphical depiction of systems via an equilateral triangle having three-components which sum to a constant. A ternary diagram has some significant advantages over other graphical representations where the actual distribution of three-components is displayed as points within an equilateral triangle. The obvious advantage of this representation is that by presenting the three-variable data in a two-dimensional plot, it is easy to analyze the correlation among the three variables simultaneously. Using a ternary diagram, Bahcall et al. (1999) proposed the `cosmic triangle' to visualize the cosmological parameters \cite{1999Sci...284.1481B}. In this cosmic triangle they  plotted the matter, curvature and cosmological constant density parameters; all these density parameters are related to one another since they follow the sum rule i.e. $\Omega_{m0}+\Omega_{k0}+\Omega_{\Lambda0}=1$. In their analysis, they showed that the constraints on the cosmological parameters using cosmic triangle representation preferred the (now standard) flat $\Lambda$CDM model.

For example, suppose we have a non-linear relation between any two independent components (say $a,~b$ and $ab$). By taking the logarithm of each component in the triad one can propose a ternary plot where, a linear relationship condition is satisfied through $\log _{10}(a)+\log _{10}(b)-\log _{10}(a b)=0$. In this context, recently, Bernal et al. (2021) propose a new cosmic triangle to visualize cosmological parameters like $H_0$, $t_U$ (age of the universe)\cite{2021PhRvD.103j3533B}.

If we have a power-law like relation between three components i.e. $a=(bc)^n$, then to reconstruct a linear relation for such power-law relation, we take logarithms on both sides and thus  can re-plot a ternary diagram between $\log_{10}(a)$, $-n\log_{10}(b)$ and $-n\log_{10}(c)$ and every point on this ternary plot satisfies the condition $\log_{10}(a)-n\log_{10}(b)-n\log_{10}(c)=0$. Thus, by knowing any two out of the three parameters, one can put constraints on the third parameter. In this work, we focus on a power-law relation between cosmological parameters and put constraints on these parameters. We show that such a representation represents  the  cosmic constraints in an intuitive and illustrative manner.

\subsection{Statistical Methods}
{In this work, we use the recent compilation of Hubble parameter measurements and Type Ia SNe measurements. We apply Gaussian process to reconstruct the Hubble parameter and its derivative using these observations. For constructing the cosmic triangle we assume a non-flat $\Lambda$CDM model as a background cosmological model. By minimizing the Chi-square, we put constraints on the parameters. }

\subsubsection{Gaussian Process}
The aim of the Gaussian Process (GP) method is to reconstruct a function from data without assuming a parametrized form of the function. For example, we have  values $H(z)$ from a set of  measurements i.e. $H(z_i)\pm \sigma_{H}$, where the value of $H(z_i)$ follows a Gaussian distribution at every point of $z_i$. Suppose we want the  value of the function at an unknown point $z^\prime$. For this a covariance or kernel function $k(z,z^\prime)$ is needed which indicates that  the value of the function at $z$ is not independent of  its value at $z^\prime$ but is correlated by $k(z,z^\prime)$.

Thus, the Gaussian Process technique is non-parametric since it is dependent on the choice of the covariance function rather than a set of  parameters in an assumed model or functional form. In general, the covariance function depends only on the distance between the points $|z-{z}^\prime|$. Here, we consider the Squared Exponential (or Gaussian) kernel function (\cite{2012JCAP...08..006M,2021EPJC...81..892O}) since this function has the characteristic of being infinitely differentiable, which is important for reconstructing a function's derivative. The Squared Exponential kernel function is 

\begin{equation}\label{eq_kernel}
    k(z, {z}^\prime)=\sigma_{f}^{2} \exp \left(-\frac{(z-{z}^\prime)^{2}}{2 \ell^{2}}\right)
\end{equation}
where, $\sigma_f$ and $\ell$ are referred as the GP hyperparameters. These basically regulate the correlation-strength of the function value and the length scale of the correlation in $z$ respectively. By minimizing a log marginal likelihood function, one can calculate the value of $\sigma_f$ and $\ell$ using the observed data. For maximization, we use flat priors for $\sigma_f$ and $\ell$ for a particular choice of kernel function.

This method can also be used to reconstruct the derivative of the function i.e. $f^\prime(z)$. While reconstructing the derivative of a function the covariance between the observational points remains unchanged.  However, we also  need a covariance between the function and its derivative as well as the  covariance between the derivatives.

Finally, the Gaussian Process for $f$ and for its derivatives $f^\prime(z)$ and $f^{\prime\prime}$ are
 
\begin{equation}\label{eq_kernel_diff}
    \begin{aligned}
    f(z) & \sim \mathcal{G P}(\mu(z), k(z, {z}^\prime)) \\
    f^{\prime}(z) & \sim \mathcal{G P}\left(\mu^{\prime}(z), \frac{\partial^{2} k(z,{z}^\prime)}{\partial z \partial {z}^\prime}\right) \\
    f^{\prime\prime}(z) & \sim \mathcal{G P}\left(\mu^{\prime\prime}(z), \frac{\partial^{4} k(z,{z}^\prime)}{\partial z^2 \partial {{z}^\prime}^2}\right)
    \end{aligned}
\end{equation}
where $\mu(z)$ is the mean value of a given function $f$.

In this analysis, we use equation (\ref{eq_kernel_diff}) to reconstruct the Hubble parameter $H(z)$ and the derivative of the Hubble parameter $H^\prime(z)$ using $32$ datapoints of Hubble parameter measurements as well as the comoving distance $d_{co}$ and the derivatives of comoving distance $d_{co}^\prime$, $d_{co}^{\prime\prime}$ using 1048 measurements of Type Ia SNe.

\subsubsection{Chi-square Analysis}
The cosmological parameters of the assumed background cosmological model are determined by maximizing the likelihood $\mathcal{L}\sim\exp\left(-\chi^{2} / 2\right)$, where Chi-square $\chi^2$ is

\begin{equation}\label{eq_chi_1}
\chi^{2}\left(\mathbf{P}\right)= \left[\mathbf{{H}_{th}}\left(z_{i} ; \mathbf{P}\right)-\mathbf{{H}_{o b s}}\left(z_{i}\right)\right]^T\operatorname{\mathbf{Cov}_{i j}}^{-1}[\mathbf{{H}_{th}}(z_j ; \mathbf{P})-\mathbf{{H}_{o b s}}(z_j)]
\end{equation}
Here, $\mathbf{P}$ represents the cosmological parameters i.e. $H_0$, $\Omega_{m0}$, $\Omega_{k0}$ and $\Omega_{\Lambda0}$, whereas $H_{th}$ and $H_{obs}$ are the  theoretical Hubble parameter and observed Hubble parameter respectively.

Here $\operatorname{Cov}_{i j}$ is the total uncertainty in the Hubble parameter which includes the effect of both statistical as well as systematic errors. Thus in our analysis, we consider the full covariance matrix $\operatorname{Cov}_{i j}$ as 

\begin{equation}\label{cov_1}
    \operatorname{Cov}_{i j}=\operatorname{Cov}_{i j}^{\text {stat }}+\operatorname{Cov}_{i j}^{\mathrm{syst}},
\end{equation}
where $\operatorname{Cov}_{i j}^{\mathrm{stat}}$  denote the contributions to the covariance due to statistical errors. The contribution due to systematic errors is given by $\operatorname{Cov}_{i j}^{\mathrm{syst}}$. To compute the systematic errors carefully we refer to the analysis carried out by Moresco et al. (2020) where, we have considered the four main contributions in the covariance matrix, i.e. initial mass function, stellar library, metallicity, and stellar population synthesis models \cite{2020ApJ...898...82M}. For a detailed discussion of the full covariance matrix please refer to Section 3.1.4 of Moresco et al. (2022)  \cite{2022arXiv220107241M}.

To remove the dependency of $H_0$ on the other cosmological parameters, and by considering a flat prior for $H_0$, we marginalize the Chi-square over $H_0$ and thus equation (\ref{eq_chi_1}) becomes

\begin{equation}
    \tilde{\chi}^{2}={A}-\frac{{B}^{2}}{{C}}+\ln \frac{{C}}{{2} \pi}
\end{equation}

Here, 
${A}=\mathbf{H_{obs}}^T(z_i) \operatorname{\mathbf{Cov}}_{i j}^{-1}\mathbf{H_{obs}}(z_j)$, ${B}=\mathbf{H_{obs}}^T(z_i)\operatorname{\mathbf{Cov}}_{i j}^{-1} \mathbf{E}(z_j;\mathbf{p})$ \& \\ 
\hspace*{1.3cm} ${C}=\mathbf{E}(z_i;\mathbf{p})^T\operatorname{\mathbf{Cov}}_{i j}^{-1}\mathbf{E}(z_j;\mathbf{p})$ 
\vspace{2mm}\\
where, \\$E(z;\Omega_{m0},\Omega_{k0})=\sqrt{\Omega_{m0}(1+z)^3+\Omega_{k0}(1+z)^2+1-\Omega_{m0}-\Omega_{k0}}$.

\subsubsection{Generation of simulated $H(z)$ data}

It seems natural that future observations will enlarge the present dataset of the Hubble parameter. To study the impact of a potentially enlarged dataset, we use a mock dataset of the Hubble parameter measurements and study its impact on the cosmic triangle and Hubble phase space portrait. To simulate the Hubble parameter data, we use the simulation method as explained by Ma and Zhang (2011) \cite{2011ApJ...730...74M}. The simulation steps are as follows:
\begin{enumerate}
    \item We choose the flat $\Lambda$CDM model as a fiducial cosmological model and adopt the best fit values of $H_0$, $\Omega_{m0}$ from \cite{Planck2018}. Using these values, we calculate the fiducial value of  Hubble parameter i.e. $H_{fid}(z)$.
    \item To update the errors of observed Hubble parameter measurements, we plot their errorbars versus redshift and based on the fact that the errors of $H(z)$ seem to increase almost linearly. 
    with respect to $z$, we identify $7$ datapoints as outliers and exclude them. 
    \item By using the remaining Hubble parameters points, to estimate the errors of the simulated data, we draw  upper and lower lines which are $\sigma^+(z)=16.577z+18.440$ and $\sigma^-(z)=7.402z+2.675$ respectively. To predict the mean error for future observations, the midline of the errors is $\sigma^0(z)=11.990z+10.558$.
    \item By assuming a Gaussian distribution of the simulated datapoints, the error of the simulated $H(z)$ is  a random number drawn from a Gaussian distribution with mean $\sigma^0(z)$ and variance $\epsilon(z)$ i.e. $\tilde{\sigma}(z)=\mathcal{N}[\sigma^0(z),\epsilon(z)]$, where we define $\epsilon(z)=\dfrac{\sigma^+(z)-\sigma^-(z)}{4}$. The parameter $\epsilon(z)$ is set so that there is a $95.4\%$ probability of $\tilde{\sigma}(z)$ lying within the strip. 
    \item The values of the simulated Hubble parameter are defined as $H_{sim}=H_{fid}+\Delta H$, where $\Delta H$ can be derived from a Gaussian distribution of the form $\mathcal{N}[0,\tilde{\sigma}(z)$].
\end{enumerate}
Using the above steps, we have generated $128$ simulated values of the Hubble parameter as well as the uncertainties in the Hubble parameter.

\section{Results} \label{results}
In this paper, our aim is to put constraints on the transition redshift $z_t$ from  Hubble parameter measurements and Type Ia SNe measurements. We reconstruct the Hubble Phase Space Portrait using Gaussian Process and put constraints on the transition redshift as well as on the equation of state parameter in a model-independent way.  Further, we propose a  cosmic triangle plot to put constraints on the  transition redshift. For this purpose, we consider a non-flat $\Lambda$CDM model as a background cosmology and put constraints on its cosmological parameters. 

\subsection{Hubble Phase Space Portrait (HPSP)}\label{hpsp_sub}
The equation for the phase space portrait is

\begin{equation}\label{eq_hpsp_main}
    \dot{H}(z)=-\dfrac{3}{2}(1+\omega)H^2(z)
\end{equation}

\begin{figure}
  \centering
  \noindent
  \resizebox{\columnwidth}{!}{
  \includegraphics[width=1.5\columnwidth]{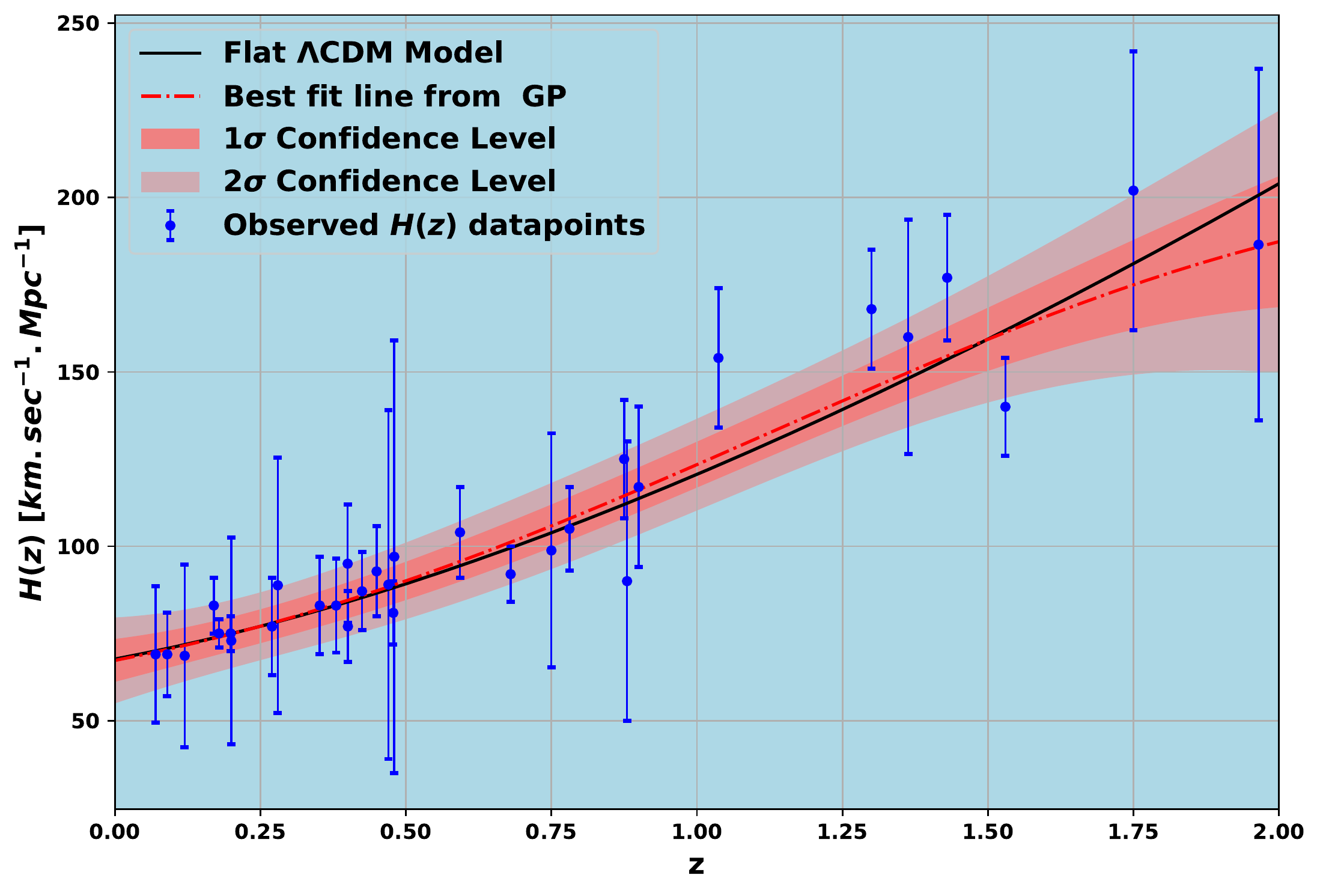}
  }
\caption{Hubble Parameter reconstruction using Gaussian Process.}
  \label{fig_gp_H}
\end{figure}

To plot this phase space portrait for $H$, we need to calculate the Hubble parameter $H(z)$ and its time-derivative $\dot{H}(z)$. For this, we use \textbf{GaPP}, a Python based Gaussian Process tool developed by  Seikel et al. (2012) \cite{2012JCAP...06..036S}  which provides the reconstruction of $H(z)$ as shown in Figure \ref{fig_gp_H}. 

The Gaussian Process, as operated by GaPP, reconstructs not only $H(z)$ from data, but also helps us to reconstruct its derivative and the associated uncertainties. For the phase space portrait, we need to estimate the time-derivative of the Hubble parameter i.e. $\dot{H}(z)$. The relation between $H^\prime(z)$ and $\dot{H}(z)$ is

\begin{equation}
    \dot{H}(z)=-(1+z)H(z)H^\prime(z)
\end{equation}

Thus, by using the reconstructed $H^\prime(z)$, we can estimate $\dot{H}(z)$  and  uncertainties in $\dot{H}(z)$ from formula,\\
$$\sigma_{\dot{H}}^2(z)=\left(\dfrac{\partial\dot{H} }{\partial H}\right)^2\sigma_H^2+\left(\dfrac{\partial\dot{H} }{\partial H^\prime}\right)^2\sigma_{H^\prime}^2+2\left(\dfrac{\partial\dot{H} }{\partial H}\right)\left(\dfrac{\partial\dot{H} }{\partial H^\prime}\right)\sigma_{HH^\prime}$$.

\begin{figure}
    \centering
    \includegraphics[width=1\columnwidth]{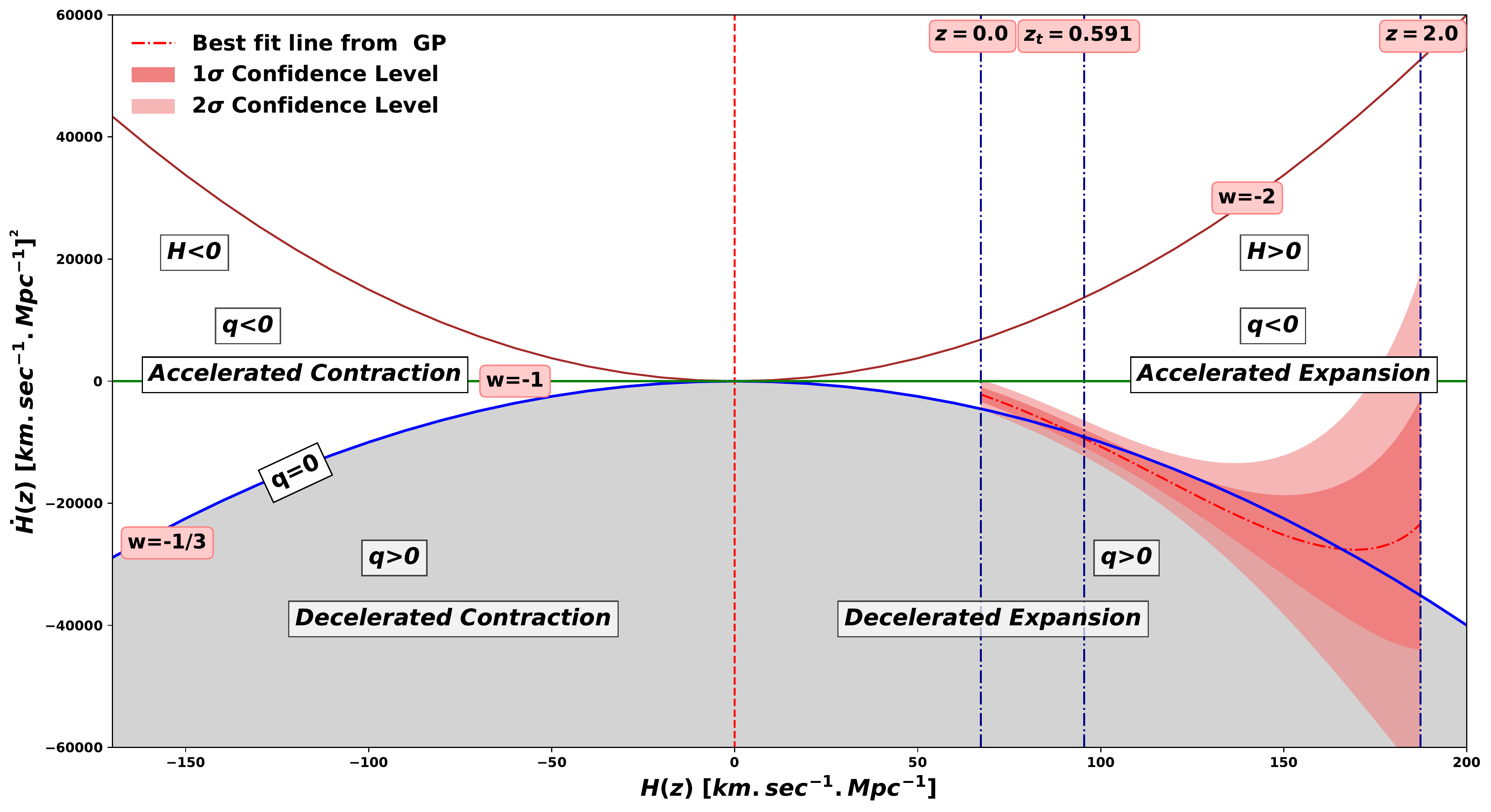}
    \caption{The Hubble Phase Space Portrait obtained by applying Gaussian process on observed $H(z)$ datapoints.}
    \label{fig_gp_Hdot_H}
\end{figure}

 Using  $H(z)$ and $H^\prime(z)$, we have drawn the  Hubble Phase Space Portrait (HPSP) between $\dot{H}(z)$ versus $H(z)$ as shown in Figure \ref{fig_gp_Hdot_H}.

In this diagram, as expected the best fit line (in red colour) is reconstructed using Gaussian process which evolves from the decelerated expansion region to the accelerated expansion region. The point where the best fit line crosses the zero- acceleration curve is generally referred to as the transition point and the redshift corresponding to it turns out to be  $z_t=0.591\pm0.332$, which is in good agreement with the recent results of Planck Collaboration et al. (2020) \cite{Planck2018}. Further, using equation (\ref{eq_hpsp_main}) and by following the reconstructed $H(z)$ and $\dot{H}(z)$, we obtain the present value of equation of state is $\omega_0=-0.677\pm0.238$.

Similarly, for the Type Ia SNe,  we estimate the $H(z)$ and $\dot{H}(z)$ by using the derivatives of $d_{co}$ as 

\begin{equation}\label{hpsp_panth}
\begin{aligned}
    H(z)&=\dfrac{c}{d_{co}^\prime}  \\
    H^\prime(z)&=-\dfrac{c}{\left(d_{co}^\prime\right)^2}\cdot d^{\prime\prime}_{co}=-H(z)\dfrac{d_{co}^{\prime\prime}}{d_{co}^\prime}\\
    \dot{H}(z)&=-(1+z)H(z)H^\prime(z)
\end{aligned}
\end{equation}

\begin{figure}
    \centering
    \includegraphics[width=1\columnwidth]{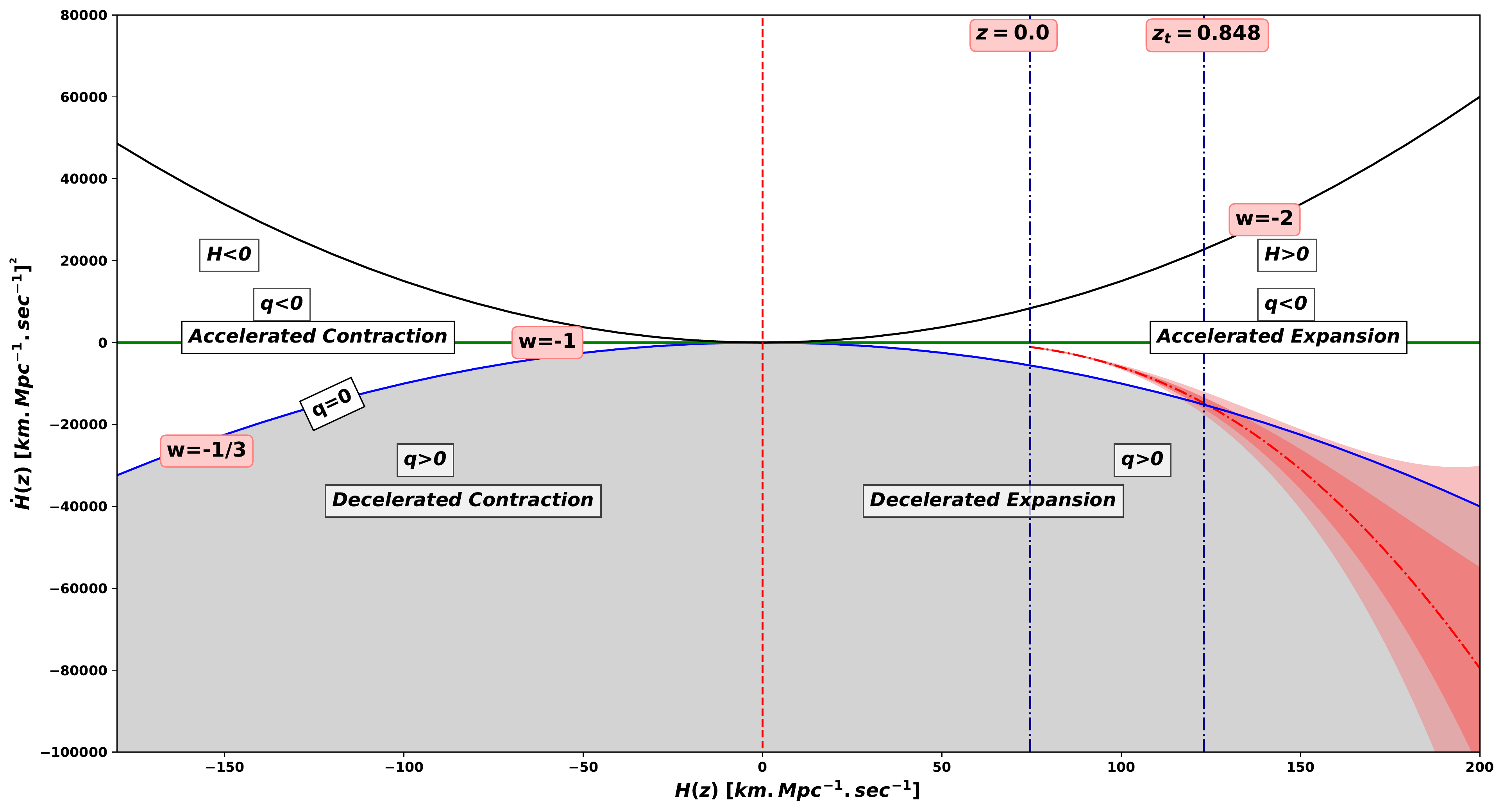}
    \caption{The Hubble Phase Space Portrait obtained by applying Gaussian process on observed Type Ia SNe datapoints.}
    \label{fig_gp_Hdot_H_panth}
\end{figure}

Thus using  Eq. (\ref{hpsp_panth}), we can draw the  Hubble Phase Space Portrait (HPSP) between $\dot{H}(z)$ versus $H(z)$ as shown in Figure \ref{fig_gp_Hdot_H_panth}.

In this case, the value of transition redshift comes out to be $z_t=0.848\pm0.177$. Further, using equation (\ref{eq_hpsp_main}), we obtain the present value of equation of state as $\omega_0=-0.870\pm0.013$.

\subsection{Results from the simulated data sets}\label{simulation_both}
So far, we have discussed the results of the Hubble Phase Space Portrait using $32$ datapoints of Hubble parameter measurements and Type Ia SNe measurements. Due to the limited number of datapoints for Hubble parameter measurements, we find  large uncertainties and hence a  wide region in both the cosmic triangle and HPSP diagrams. In order to estimate the impact of future enhancement of Hubble parameter data, we use a mock data set of the Hubble parameter measurements only and study the potential of using this type of data. In principle, one can generate a mock data set of the Type Ia SNe observations also. However, due to the large number of datapoints and small errorbars at low redshift of Type Ia SNe database, we find the value of transition redshift which has small uncertainty as compared to the value obtained from Hubble parameter measurements. Since the number of H(z) data points are less as compared to SNe data points  in the given redshift range.  Hence we simulate only the H(z) data points to  study the constraints on the transition redshift.

\begin{figure}
  \centering
  \noindent
  \resizebox{\columnwidth}{!}{
  \includegraphics[width=1\columnwidth]{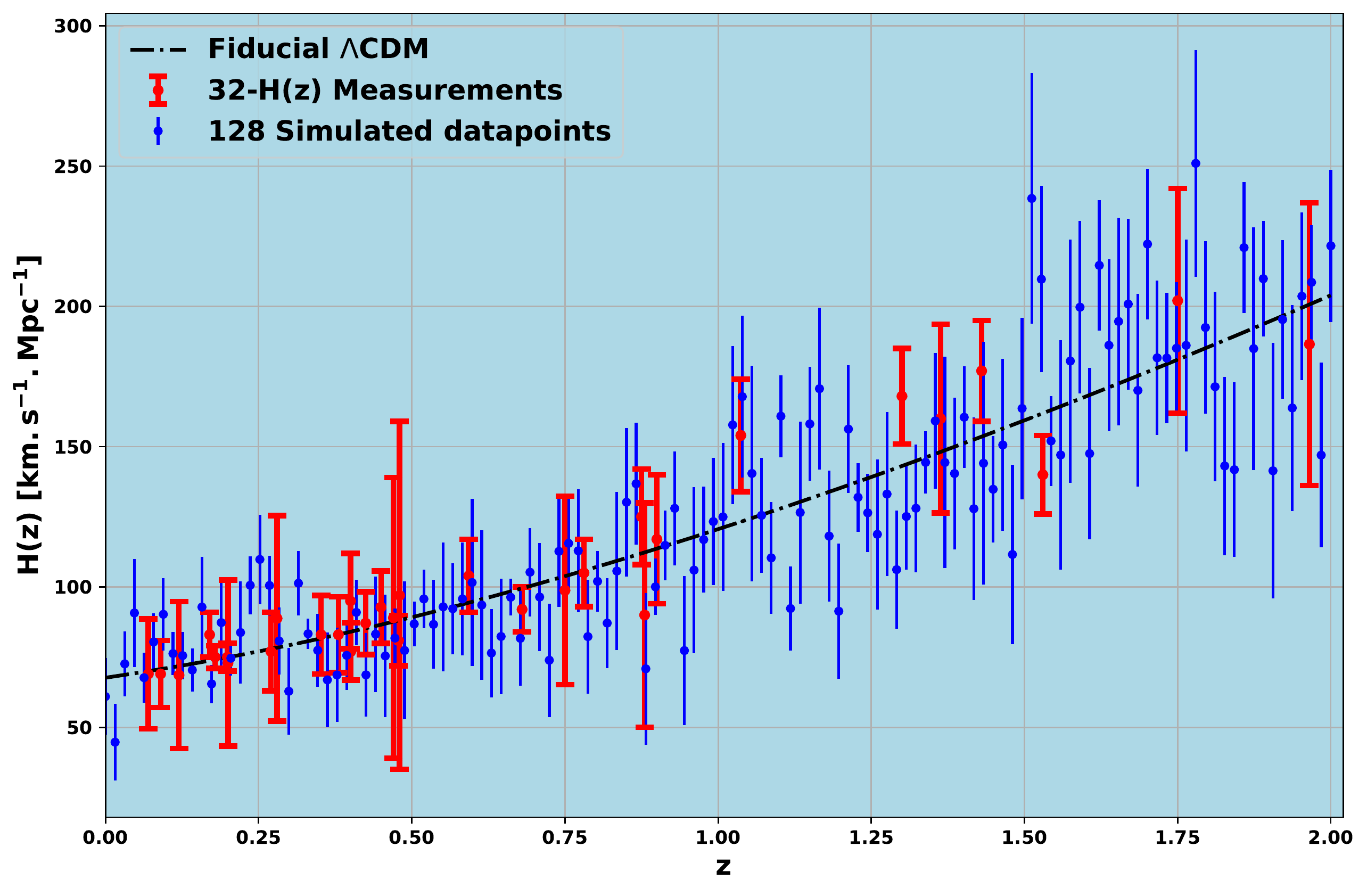}
  }
  \caption{The simulated datapoints of $H(z)$ in the redshift $0.0<z<2.0$. The blue points represent the simulated datapoints while the observed $H(z)$ are represented by the red colour. 
  }
  \label{H_simu}
\end{figure}

To simulate the Hubble parameter data we use the simulation method as described by Ma and Zhang (2011) \cite{2011ApJ...730...74M}. As explained in Section \ref{background}, we have generated $128$ simulated values of the Hubble parameter. The simulated datapoints are  shown in Figure \ref{H_simu} along with the observed Hubble parameter measurements. 

\begin{figure}
    \centering
    \includegraphics[width=1\columnwidth]{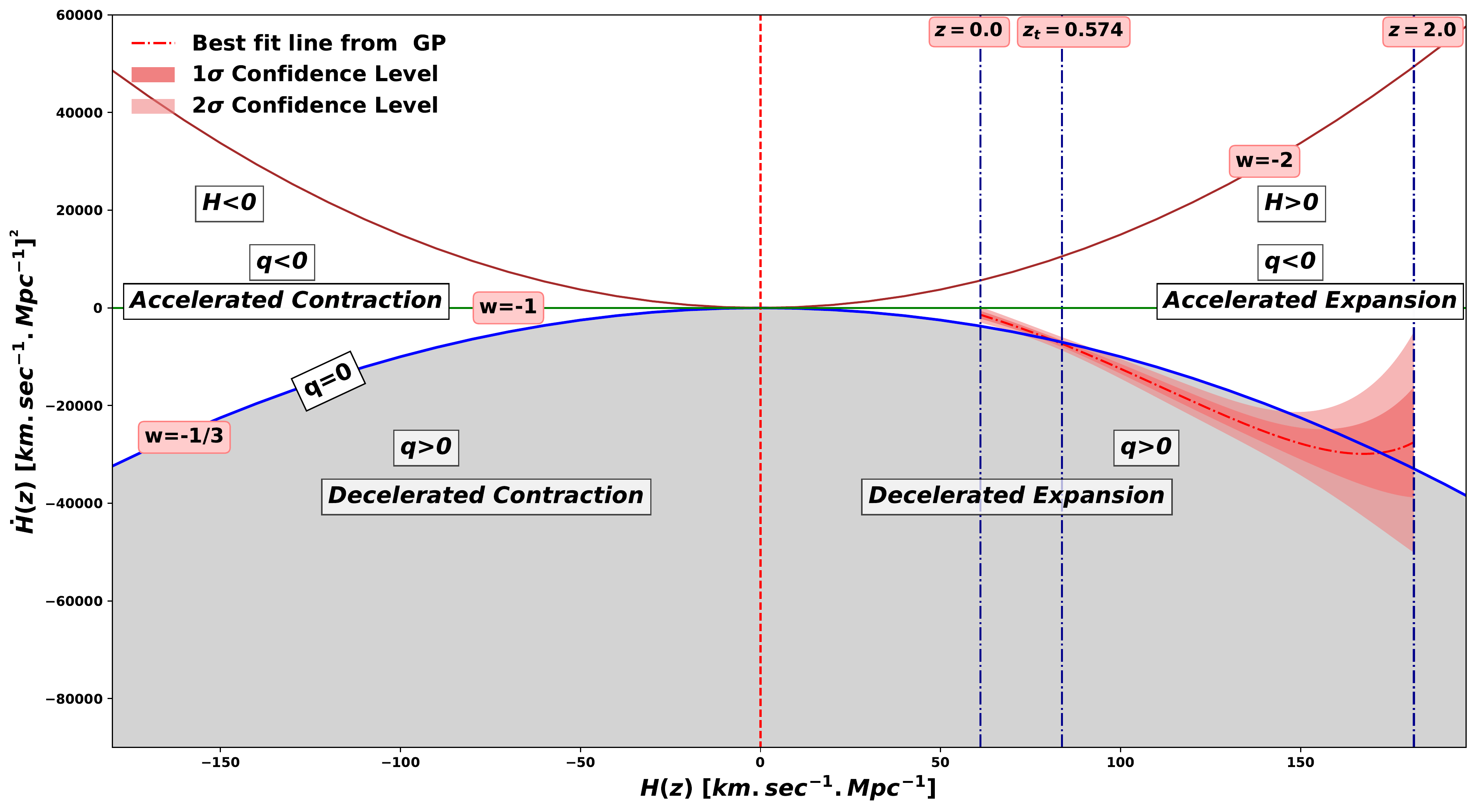}
    \caption{The Hubble Phase Space Portrait reconstructed by using 128 simulated Hubble parameter datapoints.}
    \label{hpsp_simu}
\end{figure}

Using the  simulated Hubble parameter dataset, we update the Hubble phase space diagram as shown in Figure \ref{hpsp_simu}. The best fit value of the transition redshift in this case comes out to be $z_t=0.574\pm0.193$ which is consistent with the Planck Collaboration et al. (2020) at $1\sigma$ confidence level  \cite{Planck2018} . In this case, the present value of equation of state is $\omega_0=-0.750\pm0.175$. As expected, the errors have reduced because of the increased number of datapoints.

\vspace{4mm}
For the completeness, we also perform a model-dependent analysis of estimating transition redshift using Cosmic Triangle technique. For this we consider the non-flat $\Lambda$CDM model 

\begin{equation}\label{eq_2}
    H(z)=H_0\sqrt{\Omega_{m0}(1+z)^3+\Omega_{k0}(1+z)^2+\Omega_{\Lambda0}}
\end{equation}

where, $\Omega_{\Lambda0}=1-\Omega_{m0}-\Omega_{k0}$ and $H_0$ is the Hubble constant.\\
In this model, we are left with three free parameters i.e. $H_0$, $\Omega_{m0}$ and $\Omega_{k0}$. To reduce the number of free parameters, we marginalise over $H_0$ and using  Hubble parameter measurements we put constraints on $\Omega_{m0}$ and $\Omega_{k0}$ by minimising the Chi-square. The best fit values of the  parameters obtained from Hubble parameter measurements are given in Table \ref{tab_2}.

\begin{table}
	\renewcommand{\arraystretch}{1.5}
 	\centering
    \tbl{ The best fit values of $\Omega_{m0}$, $\Omega_{k0}$ and $\Omega_{\Lambda0}$ with $1\sigma$ confidence level obtained by using 32 $H(z)$ measurements.}
   { \begin{tabular}{@{}cccccccc}
	\hline
		Parameter & Best fit value [$1\sigma$ C.L.]   \\ 	\hline
		$\Omega_{m0}$ & $0.380^{ +0.318}_{ -0.375}$ \\ 
		$\Omega_{k0}$ & $-0.186^{ +1.137}_{ -0.864}$ \\ 
		$\Omega_{\Lambda0}$ & $0.806^{ +0.544}_{ -0.806}$ \\ 
		\hline
	\end{tabular}
	\label{tab_2}}
\end{table}

The best fit value of $\Omega_{k0}=-0.186^{ +1.137}_{ -0.864}$ as shown in Table \ref{tab_2} suggests a spatially flat universe  at $1\sigma$ confidence level which is in good agreement with Planck Collaboration et al. (2020) \cite{Planck2018}.

\begin{figure}
  \centering
  \includegraphics[width=0.7\columnwidth]{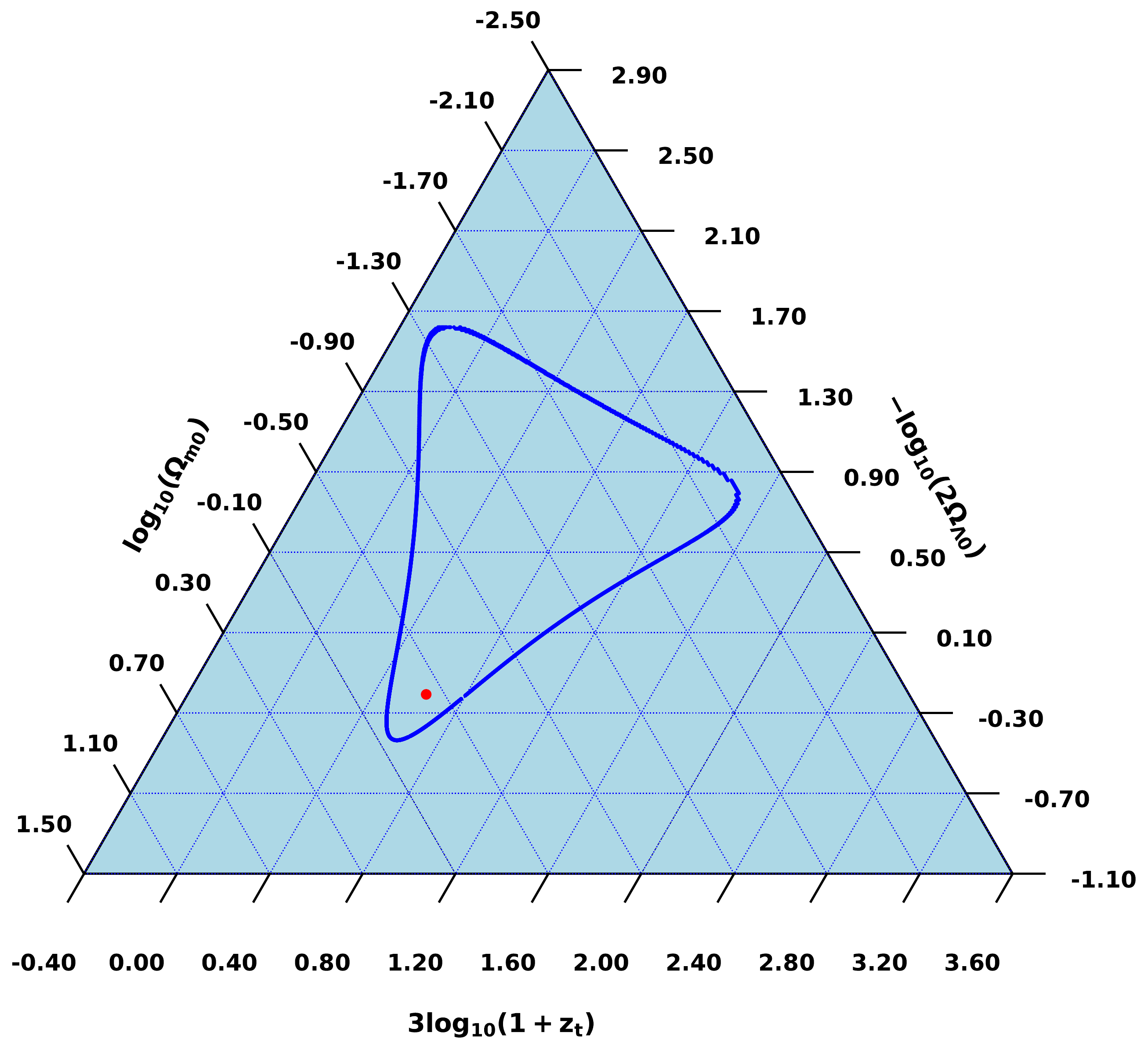}
  \caption{The cosmic triangle plot among  $\log_{10}(\Omega_{m0}),~-\log_{10}(2\Omega_{\Lambda0})$ and $3\log_{10}(1+z_t)$ with their $1\sigma$ confidence level, where the red dot denotes the best fit value of the parameters. 
  }
  \label{fig_3}
\end{figure}
 
For the transition epoch, we equate the deceleration parameter $q(z)$ to zero at redshift $z_t$. Then using the Friedmann equation, the value of the transition redshift in terms of cosmological parameters is given by
 
 \begin{equation}\label{eq_3}
     z_t=\left(\dfrac{2\Omega_{\Lambda0}}{\Omega_{m0}}\right)^{1/3}-1, ~~~~\text{where},~~\Omega_{m0}+\Omega_{k0}+\Omega_{\Lambda0}=1
 \end{equation}

Thus using the constraints of $\Omega_{m0}$, $\Omega_{k0}$ and $\Omega_{\Lambda0}$, the best fit value of transition redshift comes out to be $z_t=0.619^{+0.580}_{-0.758}$. To plot a triangle plot we require a linear relation among its three components.Taking the  logarithm of equation (\ref{eq_3}):
 
 \begin{equation}\label{eq_4}
     \log_{10}(\Omega_{m0})-\log_{10}(2\Omega_{\Lambda0})+3\log_{10}(1+z_t)=0
 \end{equation}

Using this linear relation, we reconstruct a triangle plot among $\log_{10}(\Omega_{m0}),~\log_{10}(2\Omega_{\Lambda0})$ and $3\log_{10}(1+z_t)$ as shown in Figure \ref{fig_3}.

\begin{figure}
  \centering
  \includegraphics[width=0.5\columnwidth]{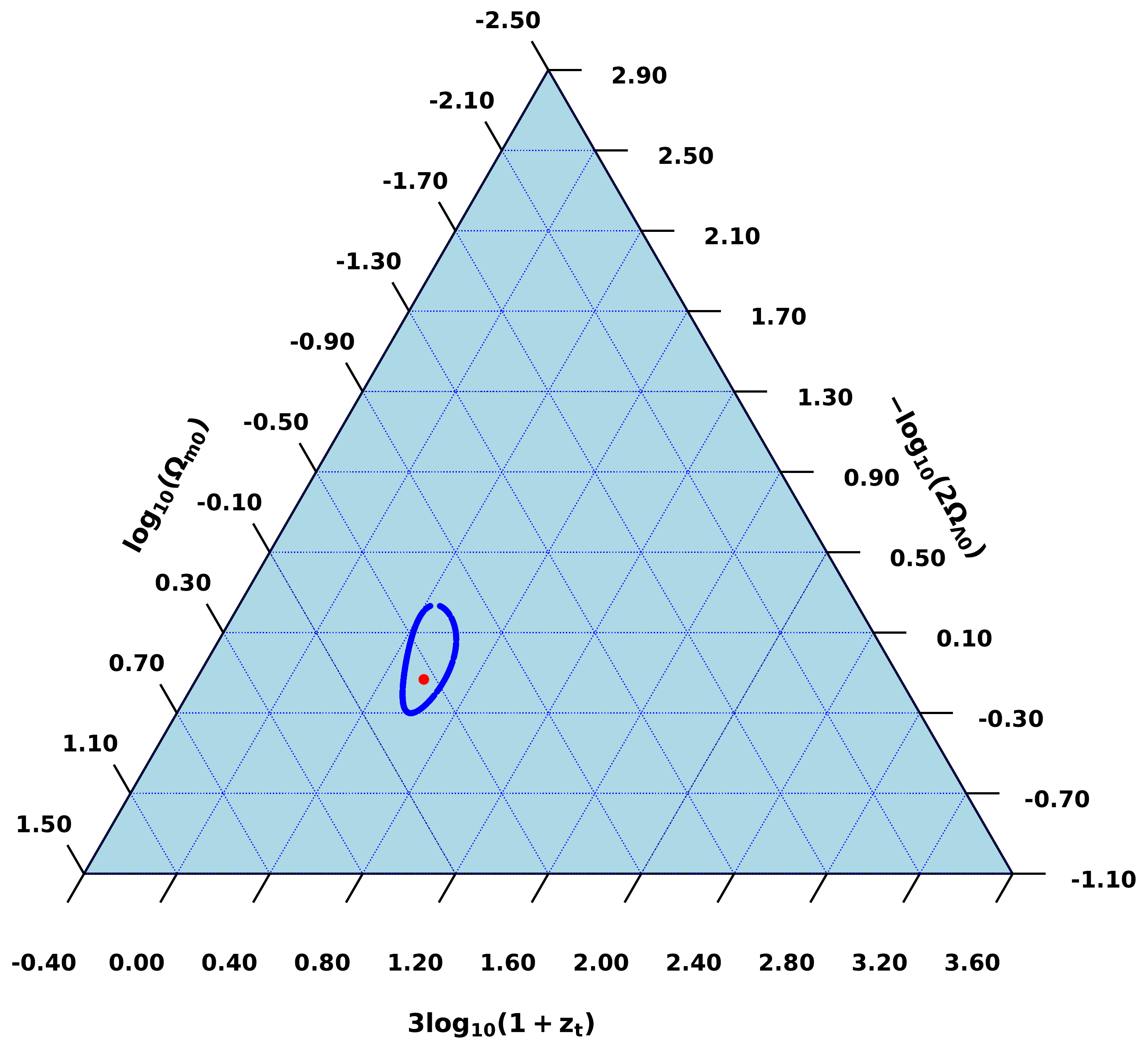}
  \caption{The cosmic triangle diagrams using $128$ simulated Hubble parameter datapoints, where the red dot denotes the best fit value of the parameters.}
  \label{ternary_simu}
\end{figure}

We re-plot the cosmic triangle diagram as shown in Figure \ref{ternary_simu} by using $128$ simulated datapoints of Hubble parameter. We find the transition redshift as $z_t=0.561^{+0.332}_{-0.379}$ and as expected the contour has shrunk substantially.

The summary of different values of transition redshift obtained from different surveys is shown in Table \ref{zt_tension}.

\begin{table}
\centering
\tbl{Constraints on $z_t$ from different cosmological observations.}
{\begin{tabular}{lccc}
\hline
Dataset & Methodology & Best fit value of $z_t$ &Reference\\
\hline
BAO, CMB, SNe & Kinematic: $\omega(z)$ &  ${\displaystyle \sim }$0.7 & \cite{2014JCAP...10..017M}\vspace{1.3mm}\\

Age of Galaxies, Strong Lensing, SNe & Kinematic: $q(z)$ & <1 & \cite{2015JCAP...12..045R}\vspace{1.3mm}\\

H(z): CC & $\Lambda$CDM Model & $0.64^{+0.11}_{-0.07}$ & \cite{Moresco_2016}\vspace{1.3mm}\\

H(z): CC+BAO & $\Lambda$CDM Model & $0.72^{+0.05}_{-0.05}$ & \cite{2017ApJ...835...26F}\vspace{1.3mm}\\

H(z): CC+BAO, SNe & Kinematic Quadratic $H(z)$ & $0.870^{+0.063}_{-0.063}$ & \cite{2018JCAP...05..073J}\vspace{1.3mm}\\

Planck-2018 & $\Lambda$CDM Model & $0.632^{+0.018}_{-0.018}$ & \cite{Planck2018}\vspace{1.3mm}\\

SNe: Pantheon & Gaussian Process & $0.683^{+0.110}_{-0.082}$ & \cite{2019ChPhC..43g5101L}\vspace{1.3mm}\\

H(z): CC+BAO & Gaussian Process & $0.59^{+0.12}_{-0.11}$ & \cite{2020JCAP...04..053J}\vspace{1.3mm}\\

H(z): CC+BAO, SNe & $\Lambda$CDM Model & $0.69^{+0.25}_{-0.25}$ & \cite{velasquez2020observational}\vspace{1.3mm}\\

H(z): CC+BAO & Gaussian Process & $0.637^{+0.165}_{-0.175}$ & \cite{2022BrJPh..52..115V}\vspace{1.3mm}\\

\textbf{Updated H(z): CC} & \textbf{Observed HPSP} & $\mathbf{0.591^{+0.332}_{-0.332}}$ & \textbf{This work} \vspace{1.3mm}\\

\textbf{Simulated H(z): CC}  & \textbf{Simulated HPSP} & $\mathbf{0.574^{+0.193}_{-0.193}}$ & \textbf{This work}\vspace{1.3mm}\\

\textbf{Updated H(z): CC} & \textbf{Observed Cosmic Triangle} & $\mathbf{0.619^{+0.580}_{-0.758}}$ & \textbf{This work} \vspace{1.3mm}\\

\textbf{Simulated H(z): CC}  & \textbf{Simulated Cosmic Triangle} & $\mathbf{0.561^{+0.332}_{-0.379}}$ & \textbf{This work}\vspace{1.3mm}\\

\textbf{SNe: Pantheon} & \textbf{Observed HPSP} & $\mathbf{0.848^{+0.117}_{-0.117}}$ & \textbf{This work} \vspace{1.3mm}\\

\hline

\bottomrule
\end{tabular}\label{zt_tension}}
\end{table}

\section{Discussion and Conclusions} \label{discussion}

That the universe is in a phase of accelerated expansion is now well established. Given this, it is an interesting exercise to determine the epoch of the transition from a decelerated phase to the accelerated phase. In this work,
we use the Pantheon database and recent data for the Hubble parameter and full covariance matrix for systematic uncertainties to put constraints on the transition redshift both in a model-independent and model-dependent way. The main conclusions are as follows:

We consider the observed value of the Hubble parameter measurements and Type Ia SNe observations to draw a phase space $(H,~\dot{H})$ portrait. Our main conclusions are listed below-

\begin{enumerate}
    \item We emphasize that the analysis with HPSP provides a way to estimate the transition redshift as well as the current value of the equation of state parameter without solving any equations for these cosmological parameters.
    
    \item In the HPSP, the reconstructed best fit line lies in the matter-dominated era  at higher redshift, which indicates that  the universe was in decelerated phase. Further, as expected, the best fit line crosses the transition boundary from deceleration  to acceleration at  a redshift, $z_t$. This characteristic of HPSP is in  agreement with the standard model of cosmology. The reconstructed best fit line in HPSP is in agreement with the   thermal history of the universe and can be used to  estimate the transition redshift. 
    The result obtained using HPSP is also in agreement with the results obtained from different observables as listed in Table \ref{zt_tension}. 
    
    \item From HPSP, the constraint obtained on the  present value of the equation of state parameter $(\omega_0)$ shows agreement with the Planck Collaboration et al. (2020) \cite{Planck2018}.
    
    \item Using the mock datapoints of Hubble parameter measurements, we reconstruct the HPSP and estimate the transition redshift value. This result is highly consistent with the results obtained from 32 datapoints of Hubble parameter measurements. 

\end{enumerate}

Finally, in order to check the value of the transition redshift by using model-dependent Cosmic Triangle technique, 
we consider a non-flat $\Lambda$CDM model as a background cosmological model. By marginalizing over $H_0$ and minimizing the Chi-square, we put constraints on cosmological parameters i.e. $z_t$, $\Omega_{m0}$, and $\Omega_{\Lambda0}$. Further, using $z_t$, $\Omega_{m0}$, and $\Omega_{\Lambda0}$, we reconstruct the cosmic triangle which allows us to visualize the confidence region in an intuitive and illustrative way. The best fit value of the cosmic curvature parameter $(\Omega_{k0})$ suggests an closed universe but a spatially flat universe can be  accommodated at $1\sigma$ confidence level. Thus our results are compatible with the Planck Collaboration et al. (2020) at $1\sigma$ confidence level \cite{Planck2018}. Using an updated cosmic triangle plot among $\log_{10}(\Omega_{m0})$, $-\log_{10}(2\Omega_{\Lambda0})$ and $3\log_{10}(1+z_t)$ we put constraints on the transition redshift, which is in agreement with the the result obtained from the different observables as listed in Table \ref{zt_tension}. We generate a mock dataset of $128$ datapoints by adopting a fiducial flat $\Lambda$CDM model. Using this simulated dataset, we find that the results are  consistent with the Planck Collaboration et al. (2020) \cite{Planck2018}. As expected, with the increase in the number of datapoints, the confidence region between these parameters reduces substantially.  For the completeness, we have performed the analysis of Gaussian process by using \textbf{six} different kernel functions. The value of the transition redshift obtained from the HPSP diagram with different kernel functions is tabulated in Table  \ref{tab_1}.

\begin{table}[h]
	\centering
	\renewcommand{\arraystretch}{2}
	\begin{tabular}[b]{| c | c|}\hline
		Kernel Function & Best Fit Value of $z_t$ [68\% CL]\\ \hline \hline
		\textbf{Square Exponential} & $\mathbf{0.591\pm0.332}$\\ \hline
		Mat\'ern 3/2 & $0.574\pm0.766$ \\ \hline
		Mat\'ern 5/2 & $0.578\pm0.433$ \\ \hline
		Mat\'ern 7/2 & $0.581\pm0.380$\\ \hline
		Mat\'ern 9/2 & $0.588\pm0.361$\\ \hline
		Cauchy &     $0.571\pm0.377$\\ \hline
	\end{tabular}
	\caption{ The best fit value of $z_t$ from Hubble Phase Space Portrait Diagram with six different GP kernel functions.}
	\label{tab_1}
\end{table}

From Table \ref{tab_1}, we can conclude that the results are not very sensitive to the choice of the kernel function.

\section*{Acknowledgements}
 We would like to thank the referee for making some very valuable suggestions. Authors are grateful to Michele Moresco and Eoin \'{O} Colg\'{a}in for their constructive and useful suggestions that helped to improve the work. Darshan thanks Jose Luis Bernal for valuable discussion. Darshan is supported by an INSPIRE Fellowship under the reference number: IF180293 [SRF], DST India. Darshan acknowledges the facilities provided by the IUCAA Centre for Astronomy Research and Development (ICARD), University of Delhi. This research work made use of the free Python packages \textsc{Numpy}  \cite{2020Natur.585..357H}, \textsc{Matplotlib} \cite{2007CSE.....9...90H}, \textsc{GaPP} \cite{2012JCAP...06..036S}, and \textsc{Python-Ternary} \cite{2015zndo.....34938H}.

\bibliographystyle{spphys}
\bibliography{main.bib} 

\end{document}